\documentclass[prl,twocolumn,amsmath,amssymb,showpacs,showkeys]{revtex4}

\usepackage{dcolumn}
\usepackage{bm}

\newcommand{\beq}{\begin{eqnarray}}
\newcommand{\eeq}{\end{eqnarray}}
\newcommand{\al}{\alpha}
\newcommand{\om}{\omega}

\newcommand{\lan}{\langle}
\newcommand{\ran}{\rangle}
\newcommand{\hz}{\hbar\rightarrow 0}
\newcommand{\lhz}{\lim_{\hz}}

\begin{document}

\title{The No-Broadcasting Theorem and its Classical Counterpart}

\author{Amir Kalev}
\email{amirk@techunix.technion.ac.il}
 \affiliation{Department of Physics, Technion-Israel Institute of Technology, Haifa 32000,
Israel.}
\author{Itay Hen}%
 \email{itayhe@post.tau.ac.il}
\affiliation{%
Raymond and Beverly Sackler School of Physics and Astronomy,
Tel-Aviv University, Tel-Aviv 69978,
Israel.
}%
\begin{abstract}
Although it is widely accepted that `no-broadcasting' -- the
nonclonability of quantum information -- is a fundamental
principle of quantum mechanics, an impossibility theorem for the
broadcasting of general density matrices has not yet
been formulated. In this paper, we present a general
proof for the no-broadcasting theorem, which applies to arbitrary density matrices. The proof
relies on entropic considerations, and as such can also be directly linked
to its classical counterpart, which applies to probabilistic distributions of statistical ensembles.
\end{abstract}


\maketitle
Concepts from quantum information theory have been shown to
provide new insights into profound topics relating to fundamental
features of quantum mechanics, such as the uncertainty principle
\cite{Uncer}, interference \cite{Inter}, entanglement \cite{Ent},
and the connection to the Second Law of Thermodynamics
\cite{QMSM}. A hallmark of quantum mechanics is that quantum
information cannot be cloned \cite{noClo1,noCloMix,noCloMixGen}. The enormous impact of this theorem,
which was called the `no-broadcasting' theorem, is reflected by several studies that
focus on various aspects of the nonclonability of quantum
information \cite{QINC1,QINC2,QINC4,QINC5,QINC6}.

Although it is widely accepted that no-broadcasting is a
fundamental principle of quantum mechanics, an impossibility theorem for the broadcasting of
general arbitrary (i.e., finite- as well as infinite-dimensional) density matrices has not yet been
formulated. In the literature two separate proofs for
no-broadcasting are to be found, one applies only to pure states
(the no-cloning theorem) \cite{noClo1} while the other applies
only to invertible density matrices \cite{noCloMix}. These two
classes of states exclude each other, and hence, trivially, none
of the two proofs is derivable from one another. (Although for
the finite-dimensional case, a generalization for non-invertible  density matrices
exists \cite{noCloMixGen}.)

In this paper we present a general proof for the `no-broadcasting' theorem
which applies to arbitrary density matrices.
Our proof relies on fundamental principles from information theory,
mainly on entropic considerations. As such,
it also enables us to directly link the theorem
to its classical analogue which applies to probabilistic distributions of statistical ensembles \cite{ClaNoClo}.

A general broadcasting machine consists of a source system whose
unknown state, $\sigma$, is to be broadcast, a target system onto
which the source state should be copied, and an auxiliary system, or a `machine',
which interacts unitarily with the source and target systems.
Labeling the three subsystems by subscripts $s$, $t$, and $m$
respectively, the broadcasting process then reads:
\begin{equation} \label{eq:rhoClone}
\rho^{in}= \sigma_s \otimes \tau_t \otimes \Sigma_m \to
\rho^{out}
\end{equation}
where the final state
$\rho^{out}$ obeys
\beq \label{eq:rhoCloneConds}
Tr_{t,m}[\rho^{out}]=Tr_{s,m}[\rho^{out}]=\sigma
\, ,
\eeq
where $Tr_{t(s),m}$ denote partial traces over the target (source) and auxiliary systems.
In what follows we  show that no unitary (quantum
mechanical) transformation which performs process
(\ref{eq:rhoClone}) exists for arbitrary
source states.

Our proof is based on the concept of relative entropy.
The relative entropy of a state $\rho_1$ with
respect to another $\rho_2$ \cite{RelDef},
\beq
S(\rho_1|\rho_2)= Tr[\rho_1 \left(\log\rho_1-\log\rho_2\right)]
\eeq
is a measure of the ``closeness'' between the
two. For some pairs
of states (``perfectly distinguishable'' ones) the relative entropy is ill-defined.
This happens if (and only if) \hbox{$\ker(\rho_2) \subseteq \ker(\rho_1)$},
yielding \hbox{$S(\rho_1 | \rho_2)=\infty$}  \cite{IllRel}.
For what follows we consider only the case \hbox{$S(\rho_1 | \rho_2)<\infty$},
and address in detail the problematic issues which may arise from this ill-definiteness, later on.
\par
One important property of relative entropy
is that it is invariant under dynamical changes.
The evolution of a general quantum system represented by a density
operator $\rho$ is given by \hbox{$\rho(t) = U(t) \rho(0) U^{\dag}(t)$}
where $U(t)$ may be any unitary operator. Since
the relative entropy is defined by a trace operation, it is easy to check that it is conserved
under time evolution, namely:
\begin{equation} \label{eq:EntDistT}
S(\rho_1(t)|\rho_2(t)) = S(\rho_1(0)|\rho_2(0)) \, .
\end{equation}
Let us now consider two general broadcasting processes (\ref{eq:rhoClone}), whose initial states are given by
\hbox{$\rho_i^{in} = \sigma_{i,s} \otimes \tau_t \otimes \Sigma_m$},
where $i=1,2$, $\sigma_{i}$ are arbitrary
density matrices, and the initial states of the target and auxiliary systems, $\tau$ and $\Sigma$,
are the same for both processes.
The relative entropy of the two states is
\begin{widetext}
\begin{eqnarray}
S\left(\rho_1^{in}| \rho_2^{in}\right) &=& Tr\bigg[ \sigma_{1,s}
\otimes \tau_t \otimes \Sigma_m
 \bigg( \log\sigma_{1,s} \oplus \log\tau_t \oplus \log\Sigma_m
 -\log\sigma_{2,s} \oplus \log\tau_t \oplus \log\Sigma_m\bigg)\bigg]    \\\nonumber
&=& Tr_{s}\left[\sigma_{1,s}
 ( \log\sigma_{1,s}-\log\sigma_{2,s} )\right]  Tr_{m,t}\left[\tau_t \otimes
\Sigma_m\right] =
 Tr_s[\sigma_{1,s} (\log \sigma_{1,s}-\log\sigma_{2,s} )]  =
 S(\sigma_{1}|\sigma_{2})\, .
\end{eqnarray}\end{widetext}
That is, the relative entropy of the two initial states  is exclusively given by the relative entropy of
the two source systems. Using this and the conservation of
relative entropy in time (\ref{eq:EntDistT}), it is clear that
\begin{equation}\label{eq:EntFinIn}
S(\sigma_{1}|\sigma_{2})=S(\rho_1^{in}|\rho_2^{in})=S(\rho_1^{out}|\rho_2^{out})\,.
\end{equation}
The relative entropy of the final states of any two broadcasting processes
is equal to the relative entropy of the sources prior to copying.
\par
We now proceed to show that Eq. (\ref{eq:EntFinIn}) is violated for broadcasting processes. To do this, we invoke
the theorem of monotonicity of relative entropy \cite{EntMon1} which reads:
\begin{equation} \label{eq:inequality}
S(\rho_{1,AB}|\rho_{2,AB})\geq
S(\rho_{1,B}|\rho_{2,B}) \, ,
\end{equation}
where $\rho_{1,AB}$ and $\rho_{2,AB}$ are two density operators of a composite system
$AB$, whereas $\rho_{1,B}$ and $\rho_{2,B}$ denote the corresponding
density operators of a subsystem $B$. The equality holds if and only if the condition
\begin{equation} \label{eq:Cond}
\log \rho_{1,AB} - \log \rho_{2,AB} =I_{A}  \otimes
(\log \rho_{1,B} - \log \rho_{2,B}) \,,
\end{equation}
evaluated after a restriction to the support of $\rho_{2, AB}$, is satisfied,
and $I_A$ denotes the identity operator of subsystem $A$ \cite{MonEntEq}.
Intuitively, Eq. (\ref{eq:Cond}) means that ignoring
part of two physical systems reduces the `distance' between them,
unless the ignored part contains no information at all.
Using (\ref{eq:inequality}), we can establish a lower bound
for the relative entropy of the two final states $\rho_i^{out}$.
The monotonicity inequality (\ref{eq:inequality}) implies that
the final states fulfill
\begin{eqnarray} \label{eq:lowerBoundS}
S(\rho^{out}_1|\rho^{out}_2) &\geq & S(\rho^{out}_{1,k}|\rho^{out}_{2,k})\,,
\end{eqnarray}
for $k=s,t$ where $\rho^{out}_{i,s(t)}$ denotes $Tr_{t(s),m}[\rho^{out}_i]$.
According to Eq. (\ref{eq:Cond}),
the equality in (\ref{eq:lowerBoundS}) holds if and only if
the equalities
\begin{eqnarray} \label{eq:EqualityBound}
&&\log \rho^{out}_1 - \log \rho^{out}_2
=(\log \rho_{1,s}^{out} - \log \rho_{2,s}^{out}) \otimes
I_t\otimes I_m   \nonumber\\
&&=   I_s \otimes (\log \rho_{1,t}^{out}-\log
\rho_{2,t}^{out})\otimes I_m \,,
\end{eqnarray}
evaluated on the support of $\rho_2^{out}$, are satisfied.
Under broadcasting, Eq. (\ref{eq:EqualityBound}) thus reads
\beq
&&\log \rho^{out}_1 - \log \rho^{out}_2
=(\log \sigma_{1,s} - \log \sigma_{2,s}) \otimes
I_t\otimes I_m   \nonumber\\
&&=   I_s \otimes (\log \sigma_{1,t}-\log
\sigma_{2,t})\otimes I_m \,.
\eeq
The above condition, however, is satisfied only if $\sigma_{1}$ and $\sigma_{2}$ are diagonal,
reflecting the fact that a realization of a broadcasting machine may be possible only provided
that all its input states are mutually commuting and the basis in which they are diagonal is known \cite{com1}.
For any two non-commuting arbitrary states
the inequalities in (\ref{eq:lowerBoundS}) are strict, that is,
\begin{equation} \label{eq:StrictlowerBound}
S(\rho^{out}_1|\rho^{out}_2) >
S(\sigma_{1}|\sigma_{2}) \,,
\end{equation}
in contradiction with equality (\ref{eq:EntFinIn}).
We have therefore shown that under broadcasting, the monotonicity of relative entropy
is in conflict with quantum dynamics, rendering universal broadcasting impossible.
\par
To complete our proof, let us consider the case of $S(\rho_1 | \rho_2)= \infty$,
and show that the no-broadcasting theorem may be extended
to this case as well \cite{Donald}.
This is done using a proof by contradiction.
Let us first assume the existence of a machine capable of broadcasting states with infinite relative entropy,
and consider two such non-commuting states $\sigma_1$ and $\sigma_2$ for
which $S(\sigma_1 | \sigma_2)= \infty$.
Due to the linearity of the broadcasting procedure (containing only unitary operations and partial traces),
it immediately follows that the machine is also capable of broadcasting the mixture
$\sigma_{\textrm{mix}}=\lambda \sigma_1 + (1-\lambda) \sigma_2$
for any $0 < \lambda < 1$.
However, our main proof rules out the existence of a machine which
broadcasts both $\sigma_1$ and $\sigma_{\textrm{mix}}$,
since $S(\sigma_1 | \sigma_{\textrm{mix}})< \infty$ \cite{KerMix}.
Therefore, the existence of a machine which broadcasts both $\sigma_1$ and $\sigma_2$
is also ruled out, contradictory to our initial assumption,
and this completes the proof.

At this point, we turn to show that the proof given above
enables a direct link between the quantum theorem and its  classical
analogue \cite{ClaNoClo}.
The classical no-broadcasting theorem
states that it is impossible to broadcast classical
probability distributions with unit fidelity in a deterministic
manner once infinitely-narrow distributions (i.e.,
delta-function distributions) are excluded;
assuming Liouville evolution for the broadcasting
process, the monotonicity of the classical relative entropy
(the Kullback-Leibler information distance)
between two classical probability distributions
$P_1(x,p,t)$ and $P_2(x,p,t)$, defined by \cite{Kull}
\beq
\mathcal{K}(P_1,P_2) = \int dx dp P_1 \left( \log P_1 - \log P_2 \right)\,,
\eeq
is in conflict with broadcasting. (We note here however, that approximate classical broadcasting machines
may in principle be realized with any desired degree of accuracy \cite{Braun}.)
\par
As we shall now show, the quantum no-broadcasting theorem translates in the $\hz$ limit
to its classical analogue if Hamiltonian dynamics, which is a subclass
of Liouville dynamics, is concerned.
This will be accomplished in two steps.
First, we show that for every classical probability distribution one can construct
a corresponding density matrix such that in the classical limit,
quantum relative entropy reduces to classical relative entropy.
Secondly, we show that quantum (unitary) dynamics reduces in the classical limit
to Hamiltonian dynamics under this correspondence. Even though these
two statements seem reasonable, even expected, to the best of our knowledge
they have not yet been shown explicitly.
\par
As a preliminary step,
we make a classical-quantum correspondence by assigning to each
classical probability distribution over phase space
$P(x,p)$ (we drop the time index $t$),
a quantum state according to
\beq\label{eq:RhoP}
\rho = \int dx dp P(x,p) | \al \ran \lan \al | \,,
\eeq
where $|\al\ran$ is a coherent state with
$\al = \frac1{\sqrt{2 \hbar \omega}} (\omega x +i p)$ (we shall fix $\omega=1$ in the following).
\par
This correspondence is of course the identification
of classical probability distributions with the $P$-representations \cite{GS}
of density matrices. Although the $P$-representation
is known to be problematic, being highly-singular, negative or even undefined,
we stress here that these types of states are not of our concern here,
since in our correspondence, the  $P$-distributions we consider
are {\it bona-fide} classical distributions.

First, we show that in the classical limit,
the relative entropy of two density matrices constructed from two classical statistical
distributions by (\ref{eq:RhoP}) reduces to the relative entropy of the
two  distributions, namely:
\begin{eqnarray}\label{eq:EntKul}
\lhz S(\rho_1|\rho_2)=\mathcal{K}(P_1 | P_2).
\end{eqnarray}
Expanding the logarithms appearing in the expression for the quantum relative entropy in a Taylor series
and then tracing term by term, it becomes sufficient to show that
\beq\label{eq:lim}
\lhz Tr [\rho_{1}(\rho_2/\hbar)^{n-1}]=\int dx dp \,P_1(x,p)\, P_2^{n-1}(x,p)\,,
\eeq
where $n \in \mathbb{Z}^{+}$, and the extra $\hbar$ factors appearing in (\ref{eq:lim}),
are introduced into the relative entropy
by rewriting $(\log \rho_1-\log \rho_2)$ as $(\log (\rho_1/\hbar) - \log (\rho_2/\hbar))$.
Inserting (\ref{eq:RhoP}) into the left-hand-side of Eq. (\ref{eq:lim}), we have:
\beq \label{eq:Euv}
&& \lhz Tr [\rho_{1}(\rho_2/\hbar)^{n-1}] =
 \int dx_0 dp_0  \, P_1(x_0,p_0) \\\nonumber
 &\times&
\int \left( \prod_{i=1}^{n-1} dx_i dp_i  \, P_2(x_i,p_i) \right)
\lhz \frac{\exp[-(2\hbar)^{-1}\mathbf{u}^{\dag} V\mathbf{u}]}{(2 \pi \hbar)^{n-1}}\,.
\eeq
where $\mathbf{u}^{\dagger}=(x_0,p_0,x_1,p_1,\cdots,x_{n-1},p_{n-1})$
and $V$, presented in a $(2 \times 2) \otimes (n \times n)$ block form is:
\beq\label{eq:V}
V_{(2n\times 2n)}=\left(%
\begin{array}{llllll}
1 & B& 0& \cdots &0& B^{T}
\\
B^{T} &1 &B & 0& \cdots & 0
\\
0 & B^{T}& 1 & B & 0&\vdots
\\
\vdots & 0 & \ddots & \ddots & \ddots & 0
\\
0 & \cdots & 0&B^{T}&1& B
\\
B & 0 & \cdots &0 & B^{T}& 1
\end{array}
\right)_{(n\times n)}\;, \eeq $1$ and $0$ being the $(2\times 2)$ unit and zero matrices respectively,  and
$B^{T}$ is the transpose of $B=-\frac1{2} \left(1-\sigma_y\right)$.
In order to evaluate the limit,
we note that $V$ is a normal matrix and as such it can be written in the form
$V=UDU^{\dagger}$ where $D$  is its diagonal form and
$U$ is unitary with orthonormal eigenvector basis as its columns.
Computation of these eigenvectors yields:
\beq \label{eigenV}
\mathbf{e}_{kj}=\frac{1}{\sqrt{2 n}}\left(\begin{array}{c}
(-1)^k\\ i
\end{array}\right)\otimes\left(%
\begin{array}{c}
1\\\om_j\\\om_j^2\\\vdots\\\om_j^{n-1}
\end{array}
\right)\;, \eeq
with corresponding eigenvalues
$\mu_{kj}=1-\om_j^{(-1)^k}$ where $\om_j=e^{2 \pi i j/n}$, $k=1,2$
and $j=0,...,n-1$. Noting that $\mu_{1,0}=\mu_{2,0}=0$, the term
$\mathbf{u}^\dag V\mathbf{u}$ in the exponent of (\ref{eq:Euv})
can thus be simplified to \hbox{$\mathbf{u}^\dag V\mathbf{u}=
\mathbf{v}^\dag D\mathbf{v}
=\sum_{k=1}^2\sum_{j=1}^{n-1}\mu_{kj}v_{kj}^2$}, with
$\mathbf{v}^{\dagger} \equiv \mathbf{u}^{\dagger} U$. The limit
thus becomes:
\beq\label{eq:delta} \lhz \frac{\exp[-(2
\hbar)^{-1}\mathbf{v}^\dag D\mathbf{v}]}{(2\pi \hbar)^{n-1}}
=\prod_{i=1}^{n-1} \delta(x_i-x_0)\delta(p_i-p_0)\,.
\eeq
This
completes the derivation of Eq. (\ref{eq:lim}), and thus proves
(\ref{eq:EntKul}).

Next, we turn to prove that the limit given in (\ref{eq:EntKul}) holds under time evolution.
This is achieved by showing that quantum dynamics is reduced
 to Hamiltonian  dynamics in the $\hbar \to 0$ limit,
provided an appropriate correspondence between classical and quantum systems is made. The proof is as follows.
\par
In classical mechanics, a statistical distribution $P_C(x,p)$
evolving in time (the time index $t$ is suppressed)
under some Hamiltonian $H(x,p)$ obeys the well-known Liouville equation \cite{vanKampen}.
In terms of the characteristic (Fourier transformed) function defined by
$P_C(x,p)=\int d \lambda d \mu  \, \tilde{P}_C(\lambda,\mu) e^{i (\lambda x + \mu p)}$,
and an analogous definition for $H(x,p)$,
the equation translates to
\beq \label{eq:C}
\partial_t \tilde{P}_C(\lambda,\mu)&=
&\int d \lambda' d \mu' \,
\tilde{P}_C(\lambda',\mu') \tilde{H}(\lambda-\lambda',\mu-\mu')\nonumber\\&\times& K_C(\lambda,\mu, \lambda',\mu') \;,
\eeq
with a `classical kernel' $K_C= \lambda' \mu - \lambda \mu'$.
Accordingly, a general quantum state (also written in characteristic form)
\beq \label{eq:rhoAndH}
\rho=
\int dx dp \int d \lambda d \mu \, e^{i(\lambda x + \mu p)} \tilde{P}_Q(\lambda,\mu)
 | \alpha \ran \lan \alpha | \;,
 \eeq
 whose evolution is governed by the Hamiltonian \cite{footnote}
 \beq
\hat{H}=
\frac1{2\pi \hbar} \int dx dp \int d \lambda d \mu \, e^{i(\lambda x + \mu p)} \tilde{H}(\lambda,\mu)
 | \alpha \ran \lan \alpha |
\eeq obeys the von Neumann equation $\partial_t \rho = i
\hbar^{-1} [\rho, \hat{H}]$. Expressing the equation in terms of
$\tilde{P}_Q(\lambda,\mu)$ and $\tilde{H}(\lambda,\mu)$, the equation takes
the form (\ref{eq:C}) but with a `quantum kernel'
\hbox{$K_Q=\frac{2}{\hbar} e^{ \frac{\hbar}{2} ( \lambda'(\lambda
-\lambda') + \mu' (\mu - \mu') ) } \sin \frac{\hbar}{2}(\lambda'
\mu -\mu' \lambda)$}. It is easy to check that
$\displaystyle{\lim_{\hbar \to 0} K_Q= K_C}$, thus we have shown
that {\it every} classical system may be viewed as a limiting case
of an appropriately constructed quantum system. Together with the
result of the classical limit of the relative entropy
(\ref{eq:EntKul}), the no-broadcasting theorem which states that
the monotonicity of relative entropy of two density operators
is in conflict with quantum dynamics under broadcasting,
translates in the classical limit to its classical version
\cite{ClaNoClo}, stating that the monotonicity of (classical)
relative entropy is in conflict with Hamiltonian dynamics.
\par
As with other
results from quantum mechanics that have their analogies and parallels in classical
probabilistic theories \cite{Holevo,Schu,Hor,KMR,Braun},
the classical no-broadcasting theorem can also be recovered
from its quantum version. This reduction is attributed to the fact
that both quantum and classical information theories are based on
common grounds and are described by analogous measures.
\par
We have thus shown that no-broadcasting is indeed a general principle,
originating from fundamental concepts of information theory, in particular, the monotonicity
of relative entropy.
\par We believe that this may help to gain a better
understanding of the relations between nonclonability and
reversibility properties both in quantum and in classical physics.
This proof may also provide a further clarification
on ``quantumness'' versus ``classicality'' in that context,
in particular in connection with a recent result by Walker and Braunstein \cite{Braun},
who proved the realizability of approximate classical broadcasting of statistical distributions
with any desired degree of accuracy.

We thank Venketeswara Pai, Sam Braunstein, Gilad Gour, Ady Mann,
and two anonymous referees for useful comments.


\begin{thebibliography}{100}
\bibitem{com1}
That mutually commuting states are broadcastable
provided the basis in which they are diagonal is known,
was also established in the formulation of the original no-cloning and
no-broadcasting theorems \cite{noClo1,noCloMix}.
\bibitem{KerMix}
First we note that for every vector $|x \rangle$ and a density matrix $\sigma$,
$\langle x | \sigma | x \rangle \geq 0$ with equality iff $| x \rangle$ belongs to the kernel of $\sigma$.
Considering
$\langle x | \sigma_{\textrm{mix}} | x \rangle =
\lambda \langle x | \sigma_1 | x \rangle +(1-\lambda) \langle x | \sigma_2 | x \rangle$,
we notice that the left hand side is zero iff both bracketed terms on the right hand side are zero as well,
meaning that $\ker(\sigma_{\textrm{mix}})=\ker(\sigma_1) \bigcap \ker(\sigma_2)$
and in particular $\ker (\sigma_{\textrm{mix}}) \subseteq \ker(\sigma_1)$.
\end{thebibliography}
\end{document}